\documentclass[sigconf]{acmart}
\acmConference[SE 2030]{International Workshop on Software Engineering in 2030}{November 2024}{Puerto Galinàs (Brazil)}
\AtBeginDocument{%
  \providecommand\BibTeX{{%
    \normalfont B\kern-0.5em{\scshape i\kern-0.25em b}\kern-0.8em\TeX}}}

\setcopyright{acmlicensed}
\copyrightyear{2018}
\acmYear{2018}
\acmDOI{XXXXXXX.XXXXXXX}

\acmBooktitle{Woodstock '18: ACM Symposium on Neural Gaze Detection,
 June 03--05, 2018, Woodstock, NY} 
\acmISBN{978-1-4503-XXXX-X/18/06}

\usepackage{ifthen}
\usepackage[normalem]{ulem} 
\definecolor{myGreen}{HTML}{009900}
\newboolean{showcomments}
\setboolean{showcomments}{true} 
\ifthenelse{\boolean{showcomments}}
  {

  }
		
\newcommand{\ext}[1]{}

\begin{document}


\title[Early V\&V for System Behaviour]{A Road-Map for Transferring Software Engineering methods for Model-Based Early V\&V of Behaviour to Systems Engineering}

\author{Johan Cederbladh}
\email{johan.cederbladh@mdu.se}
\orcid{0000-0003-2021-8341}
\affiliation{%
  \institution{Mälardalen University}
  \streetaddress{Universitetsplan 1}
  \city{Västerås}
  \country{Sweden}
  \postcode{72228}
}

\author{Antonio Cicchetti}
\email{antonio.cicchetti@mdu.se}
\orcid{0000-0003-0416-1787}
\affiliation{%
  \institution{Mälardalen University}
  \streetaddress{Universitetsplan 1}
  \city{Västerås}
  \country{Sweden}
  \postcode{72228}
}

\renewcommand{\shortauthors}{Cederbladh and Cicchetti}

\begin{abstract}
In this paper we discuss the growing need for system behaviour to be validated and verified (V\&V'ed) \textit{early} in model-based systems engineering. Several aspects push companies towards integration of techniques, methods, and processes that promote specific and general V\&V activities earlier to support more effective decision-making. As a result, there are incentives to introduce new technologies to remain competitive with the recently drastic changes in system complexity and heterogeneity. Performing V\&V early on in development is a means of reducing risk for later error detection while moving key activities earlier in a process. 

We present a summary of the literature on early V\&V and position existing challenges regarding potential solutions and future investigations. In particular, we reason that the software engineering community can act as a source for inspiration as many emerging technologies in the software domain are showing promise in the wider systems domain, and there already exist well formed methods for early V\&V of software behaviour in the software modelling community. We conclude the paper with a road-map for future research and development for both researchers and practitioners to further develop the concepts discussed in the paper. 
\end{abstract}




\maketitle

\section{Introduction} \label{Sec:Introduction}







The International Council On Systems Engineering (INCOSE) defines a system as follows: ``\textit{A system is an arrangement of parts or elements that together exhibit behaviour or meaning that the individual constituents do not}'', and can be either conceptual, physical, or a combination of both \cite{walden2023systems}. Modern systems are seeing increased complexity in the wider engineering landscape, partly due to the increased prevalence of software but also complexity in system functions and couplings \cite{antinyan2020revealing}. In conjunction, customers' needs are shifting toward sustainable solutions and the traditional product is being replaced by services and capabilities changing existing business models. This multi-faceted change in the landscape stresses existing practices and demands change in the current processes and technologies. Potts \textit{et al.} performed an empirical study in the Systems Engineering (SyE) community and concluded that two categories of complexity stand out among companies, namely Organisational and System behaviour aspects \cite{potts2020assaying}. In their paper they define organisational complexity as ``\textit{The number, diversity, level of support, and involvement of internal and external system stakeholders}'', while behaviour complexity is defined as ``\textit{The ability to define and predict system modes, functions, states, behaviour, performance, and missions, including degree of autonomy and the impact of the environment.''}. Indeed, this finding maps well to several emerging technologies like Cyber-Physical Systems (CPSs) \cite{baheti2011cyber}, Digital Twins (DTs) \cite{ferko2022architecting}, Systems of Systems (SoSs) \cite{maier1998architecting}, and similar domains. In fact, these technologies push for more integrated and interconnected systems, thus increasing the demands of organisational collaboration, while at the same time resulting in increased heterogeneity in the solution domain (e.g., software contra hardware) \cite{walden2023systems}. A methodology that is considered to potentially alleviate this complexity is to rely on \textit{models} as primary artefacts during development \cite{madni2018model}, where a model can be defined as an abstract representation of something real for a cognitive purpose. Starting from this approach many sub-methodologies have emerged like Model-Based Systems Engineering (MBSyE) \cite{wymore2018model}, Model-Driven Engineering (MDE) \cite{schmidt2006model}, Model-Based Engineering (MBE) \cite{liebel2018model}, etc. In the systems domain the adoption of models is progressively becoming a standard practice, and INCOSE foresee the future of SyE to be ``\textit{predominantly model-based}''\footnote{The INCOSE 2035 vision: https://www.incose.org/publications/se-vision-2035}. Modelling is seen as an essential part of SyE to not only leverage abstraction to address increasing system complexity, but also as a foundation to formalise concepts and integrate surrounding emerging technologies effectively \cite{madni2018model}. 

As the SyE discipline is founded in different standard processes the modelling activities need to properly be integrated to support and improve company efficiency; for instance, ``too much'' modelling could be detrimental \cite{cederbladh2023light}. Modelling should support processes for managing the risk that the systems to be delivered are \textit{erroneous} while keeping development efficient \cite{walden2023systems}. Therefore, modelling should support the different sub-categories related to SyE, such as systems thinking, project management, and general design approaches \cite{haberfellner2019systems}. Most of the time development happens in stages, an example is seen in the ISO 15288 reference life cycle standard for systems and software\footnote{ISO 15288 can be found at: https://www.iso.org/standard/63711.htm} and the well-known V-model \cite{walden2023systems}. These stages are in general performed in steps (with more or less iteration depending on the project at hand) and eventually lead to a delivered product design to be implemented and produced (possibly in increments). Industry best practices warn us that the cost of addressing issues in a system design increases exponentially as the development progresses. Therefore, it is of interest to detect and address any issues as early as possible. Similarly, making design decisions as early as possible is valuable since it reduces overall development time/effort \cite{walden2023systems}. In this context, performing behaviour Validation and Verification (V\&V) at early stages is required to anticipate design decisions while maintaining confidence in the system correctness. At the same time, it is a challenging goal when trying to keep the development processes efficient.

In this paper we extract challenges found in literature for early V\&V of system behaviour (referred to as early V\&V later in the document) and discuss potential future research directions. Notably, we take inspiration from the Software Engineering (SwE) discipline as the foundations for model-based development is stronger, and languages such as UML (and subsequently SysML for systems) are widely known and used \cite{cederbladh2023early}. The rest of the paper is structured as follows: Section \ref{Sec:Background} further describes early V\&V and its relation to System development. In Section \ref{Sec:Motivation} we summarise and present the current challenges while Section \ref{Sec:Approach} provides a high-level road-map for solving the challenges. Finally Section \ref{Sec:RW} presents and discusses related work while Section \ref{Sec:Conclusion} concludes the paper. 

\section{Early V\&V and its context} \label{Sec:Background}

An overarching goal of applying different system and software methodologies is to support efficiency and promote quality in development. Over-time the best practice has changed, for example from techniques such as the waterfall, to the V-model and Agile \cite{walden2023systems}. While different approaches have corresponding strengths and weaknesses, eventually they should assist practitioners in efficient and correct development. As one of the foundations of SyE is V\&V \cite{walden2023systems}, all approaches consider it as part of the overall system development. \textit{Early} as such is dependent on the process at hand and fundamentally a \textit{relative} concept. Therefore, the \textit{early stage} in SyE is difficult to strictly define \cite{cederbladh2023early}. Also, the stakeholders will not necessarily be shared between the underlying system views, especially when considering heterogeneous systems with several domains, thus leading to different definitions of what is early or not even for the same system. Nonetheless, when reviewing scientific literature some commonalities about the early stage V\&V concept can be found \cite{cederbladh2023early,ponsard2007early,carroll2016systematic}, and we formulate a set of characteristics as:

\begin{description}

\item[C1:] The threshold for acceptable uncertainty is higher compared to later development.
\item[C2:] The goal is pruning the design space rather than eliciting a singular design choice. 
\item[C3:] The properties of interest are more abstract/holistic compared to later development.
\item[C4:] The engineer prioritisation regards models/knowledge re-use over the contextual model fidelity.
\item[C5:] The effort of new model development is low.
\item[C6:] The decision-making affects a wide area of viewpoints and stakeholders.

\end{description}

At a glance, early V\&V should reduce subsequent efforts by re-using artefacts and/or knowledge from previous endeavours. Similarly, as the early stage inherently consists of high uncertainty due to the nature of development the granularity of analysis is limited. One common example is that parameters are estimated or based on other examples, such as re-using the parameterisation for a component from another project \cite{lange2018systematic}. Therefore, some margins and conservatism to make sure that the eventual analysis does not over-promise are required, as things might change in requirements and technical design based on fluctuations in customers' needs/expectations during development \cite{walden2023systems}. Correspondingly, analysis often regards more abstract Key Performance Indicators (KPIs) at early stages, which will often aim to find a set of \textit{good enough} solutions rather than the \textit{correct solution} \cite{Cederbladh2023Colab}. Another reason why the analysis strictness should be adequate is due to the multi-stakeholder concerns that should be addressed early on. Taking a high-level design decision inherently propagates to several stakeholders, and should therefore find a suitable trade-off to not restrict the design space excessively while still pruning unsatisfactory solution groups. This process is then repeated (with some alterations) until the eventual system design has been decided \cite{walden2023systems}.

To visualise the evolution of a system during development a typically used representation is the ``cone of uncertainty'' \cite{aroonvatanaporn2010reducing}. The system is developed via some well-formed process with different V\&V gates throughout the development. During the process any given KPI will naturally change as the system is being designed and decisions affect the shape of the solution. Given that early V\&V is intrinsically difficult and the certainty of results is reduced the \textit{earlier} analysis takes place, it might pose the question of why doing it at all? From the industrial perspective, it relates to market competitiveness and the potential reduction of time to market, where any improvement is a potential advantage over competition. Indeed, reducing time spent on (particularly manual) activities is valuable in itself \cite{bucchiarone2021future}. Likewise, in some domains such as Safety-Critical Systems, it can be a means of reducing potential complex and effort-heavy analysis by introducing early evaluations \cite{ponsard2007early}. 

\textit{Behaviour} is a term that has intuitive meaning in many domains, but is difficult to strictly define in a broader context. In this paper we adopt the definition given by Ackoff for behaviour in the context of systems \cite{ackoff1971towards}. Broadly, behaviour is defined as ``\textit{A system's behaviour is a system event(s) which is either necessary or sufficient for another event in that system or its environment}''. Events are intrinsically linked with change, and events for which the consequences of the change are of interest are considered a systems behaviour. Naturally, there are different kinds of classifications and descriptions for system behaviour depending on the system and its environment: state-based behaviour, goal-seeking behaviour, process-based, a combination of the previous types, or a specialised version, are few examples. Moreover, the concept of a system behaviour will depend on the system itself, and also the maturity of the system definition. So, while the motivation for early V\&V is relatively similar for different domains, the eventual techniques are instead greatly varied due to the multi-faceted context of system behaviour \cite{cederbladh2023early}. Particularly, the underlying domain plays an important role in determining what the eventual properties of interest are. Likewise, the concept of early will vary accordingly as a property will not necessarily be evaluated the same way or through the same process or even for the same reason \cite{vangheluwe2002introduction}. Nevertheless, as mentioned by the common characteristics there are some patterns found in the early V\&V techniques. There is an emphasis on re-use and boosting the analytical capabilities of corresponding notations used in early stages, such as SysML/UML \cite{nigischer2021multi}. Moreover, to combat the high uncertainty pertaining to early stages and hence enable the use of more advanced/accurate analysis methods, the required domain knowledge is typically encoded in model-transformations or the target analysis notation (such as transforming a SysML model to a simulation-specific notation) \cite{cederbladh2023early}. 

For the analysis itself, early V\&V heavily relies on simulation-based technologies \cite{cederbladh2023early}. There are several reasons for why this is the case: simulation provides a flexible analysis as inputs can be varied along with parameters; there is often some intrinsic visualisation; trade-off analysis is often included or relatively easy to attach to most simulation engines; often simulation models can be re-used from various libraries while the simulation engine can manage assumptions. Other often utilised methods include model-checking, manual inspections, and prototyping (notably for more HW-centric domains) \cite{cederbladh2023early,walden2023systems}. Typically, if there is a well understood domain other technologies or more property-specific simulation can be employed. The trade-off is that more information must be captured in the chosen analysis method, hence requiring more effort to setup and maintain it (possibly reducing general applicability). 

In the next section, we discuss the current landscape for early V\&V and challenges currently inhibiting broader adoption.

\begin{figure}
    \centering
    \includegraphics[width=0.95\linewidth]{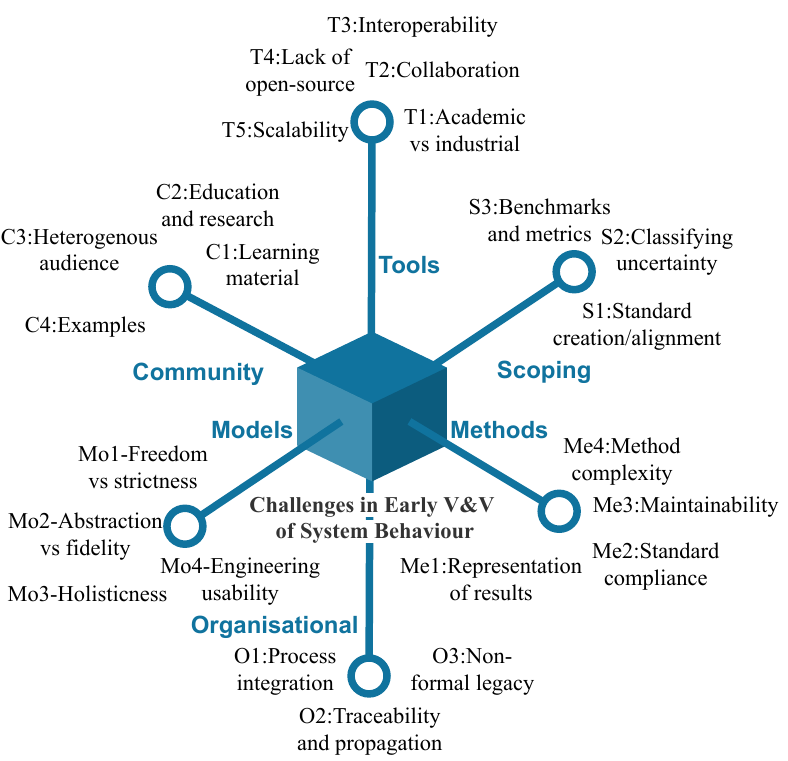}
    \caption{Challenge areas for early behaviour V\&V.}
    \label{fig:Challenges}
\end{figure}

\section{Current challenges in early V\&V} \label{Sec:Motivation}

The authors of this paper recently performed a Systematic Literature Review (SLR) in the context of early V\&V in MBSyE due to perceived gaps in literature \cite{cederbladh2023early}. One of the goals of the SLR was to identify possible barriers for industrial adoption given the relevance of early V\&V discussed so far. As a result, we elicited a set of perceived challenges as provided by the publications included in the study. Here we provide a slightly restructured list of the challenge areas in Figure \ref{fig:Challenges} together with a more detailed breakdown for each specific challenge and a corresponding discussion in the following section. At a glance, there are a few observations that can be drawn of the challenges from a software perspective: (1) Many of these challenges are also shared with SwE; (2) Some of the challenges have already been partially or fully solved in industrial segments for SwE. With those points in mind, we want to emphasise \textit{what} the main differences are between the domains, and \textit{if} there are potential knowledge transfers between both domains.  

\subsection{Models (challenges Mo1-Mo4)}

Early V\&V roughly requires two main ingredients, the analysis method and technique, and the (model) artefact(s) to use as source of information to perform the analysis. Models per definition leverage abstraction, as an enabler to reason about complex systems even at early stages. At the same time, abstraction can be a double-edged sword as it reduces overall detail in an artefact. The balance between finding the correct abstraction contra fidelity in a model is a central challenge; notably, Box and Draper \cite{box1987empirical} state that ``\textit{all models are wrong, but some are useful}''. If the model is \textit{too} abstract, it will not be useful due to the lack of relevant details. However, a model that has too many details is likewise not useful as it either takes considerable effort to be developed or cannot be fully completed at an early-stage for detailed behaviour analysis. Additionally, modelling needs to encompass a wide-audience while providing an abstract and ``easy-to-use'' language/notation. Identifying an adequate abstraction and corresponding notation/formalism is necessary but often difficult to understand before starting the development and knowing V\&V needs. Engineering usability is another issue with models, for example UML/SysML modelling notations are notoriously a challenge to learn and incorporate in the day-to-day tasks \cite{bucchiarone2021future}. Eventually, the models considered should also be holistic as the system and its environment need to be adequately captured and maintained. While languages like UML/SysML enable multi-formalism through multi-view models, it poses several challenges to adequately capture the holistic aspects and nature of systems. 


In SwE the challenges related to modelling artefacts have been discussed and reasoned for a long time \cite{bucchiarone2020grand}, and many solutions have emerged trying to tackle those. For the trade-off between model abstraction and fidelity there are initiatives like Multi-Paradigm-Modelling (MPM) reasoning that for a specific context and set of properties of interest there will be a set of suitable model notations \cite{vangheluwe2002introduction}. Regarding data, there are many efforts to standardise data and APIs, such as XML \cite{tekli2009overview}. The usability and user experience of modelling is similarly a hot topic, and efforts have been made to improve practitioners' experience \cite{law2010modelling}. Overall, the challenges experienced in MBSyE are mostly the same as for Model-Based Software Engineering (MBSwE) in terms of the models themselves, and much of the knowledge can be re-used. At this stage, what is missing in the field of MBSyE compared to MBSwE is: (1) Clear vocalisation of the challenges related to modelling; (2) Unified efforts toward model abstraction principles; (3) Harmonisation of industrial needs in practice for tools and the models themselves.

\subsection{Organisational Challenges (O1-O3)}

As MBSyE generally is expected to be applied by industrial practitioners, challenges emerge on the organisational level. Here we regard the processes involved and how they support or are supported by early V\&V. Notably, early V\&V should find an effective \textit{vertical} integration in the process, such as simulation engineers and system architects working together, however often \textit{horizontal} integration between various solution domains such as hardware and software is also required \cite{cederbladh2023early}. To effectively harmonise between these different audiences there is a need to find a common vocabulary and understanding of the technologies involved so that the V\&V results of behaviour is correctly propagated in the organisation. It might be the case that a solution \textit{can} be implemented, but \textit{should} it be, and if so, \textit{why}? Another common bottleneck is the non-formal legacy of document-based development that needs to be considered \cite{gregory2020long}. Effectively changing the organisation to consider and incorporate new technologies is a research field on its own. Nonetheless, the legacy that exists needs to be effectively incorporated into new methods and technology to support transitional phases and promote cultural and organisational changes. Lastly, at a organisational level it is very important to understand how different activities are correlated in the development, requiring strong traceability and propagation capabilities. Indeed, there needs to be a guarantee that V\&V activities of system behaviour are correct and lead to correct design decisions. Without such support, the value of early V\&V is significantly reduced and might even be counter-productive \cite{cederbladh2023early}.

In SwE the issues of how to integrate models and at what stages are much more well-understood \cite{hutchinson2011model}. Indeed MBSwE has a long history of model-transformations and code-generation. These methods explicitly emphasise source and target artefacts, and have also been increasingly tried in MBSyE, however with less success \cite{cederbladh2023light}. Partially, this is due to the fact that there is often less formalisation in the SyE context, but also because the systems as such are often only partially in software, and sometimes not at all. As a result, the approaches targeting software are difficult to directly apply, and when attempted result in ad-hoc and difficult to maintain solutions that often are seen as too ``software-centric'' by practitioners \cite{cederbladh2023light}. So, while there is an increasing need and wish to integrate the practices of MBSwE in MBSyE the differences and nuances of the context makes that challenging.

\subsection{Methods (Me1-Me4)}

While surrounding aspects are vital and pose complex challenges, at the centre of V\&V are the methods themselves. Notably, most methods or approaches available are applied ad hoc or in very specific contexts \cite{cederbladh2023early}. There is need to enable more standard-oriented approaches so that uncertainty is managed through commonly understood means and reasoning. At the same time, it is often difficult to find truly generaliseable solutions in MBSyE as there exists a broad range of scopes and it is required to embed pre-conceived knowledge in the analyses. Often this results in a need to jointly formalise implicit practitioners' heuristics, leading to frictions in abstraction and scope. A related challenge is that often the methods themselves are quite complex and leverage state-of-the-art or cutting-edge technology. As a consequence it can be difficult to interpret the results due to the abstraction gap between the artefacts used as inputs and the underlying methods used for the analysis. Additionally, most methods utilise some form of model transformation in the technical integration and these kinds of solutions can quickly become hard to maintain \cite{stevens2020maintaining}. Maintainability overall is a challenge with these methods, especially if they rely on academic tools, as there might not be a priority and/or the resources for maintaining a solution over a long period of time. Even with industrial tools maintainability is a challenge in the fast changing technology landscape. 

Many of the methods are inspired or directly taken from SwE \cite{cederbladh2023early}. As a result, there is a prevalence of model transformations and focus on the eventual delivered artefacts. In MBSyE these methods are often applicable in case-by-case implementations, but often they are not a good fit for the larger context. MBSyE emphasises the process compared to the eventual result \cite{walden2023systems}, and the required representation of said results is not necessarily the same for each domain. In general the methods of analysis should be following SyE standards and preferably be used for various certification and compliance activities during the development progression. While this is also true for many cases in SwE the underlying standards are typically not the same, and the SyE processes emphasise different concerns compared to SwE. Therefore there is a need to better align the methods and current practices with the standards that exist in the SyE community, as the mismatch creates friction between practitioners in the corresponding fields. 

\subsection{Scoping (S1-S3)}

A reoccurring issue when discussing or investigating early V\&V is that there is a gap in formalisation and definitions. Several side-effects emerge due to this, and in general discussing the topic becomes difficult. A pre-requisite for most actors and stakeholders in the SyE/MBSyE community is that any method should be compliant with standards that are used in the domain. Currently there are few attempts at aligning and creating such standards. To counteract the barrier for widely adoptable methods there is firstly the need to agree upon a common vocabulary, definitions, and attributes of the approaches. Likewise, dealing with early V\&V entails large portions of uncertainty, but in the literature there is seemingly no practical and useful classification. While uncertainty can be broadly considered as either epistemic or aleatory \cite{walden2023systems}, for early V\&V is mostly due to epistemic uncertainty, but also often due to \textit{belief} uncertainty \cite{burgueno2023dealing}. Similarly, there are few benchmarks or metrics for how to evaluate any type of technique, and the lack of empirical studies is a re-occurring issue in SyE/MBSyE \cite{henderson2021value,campo2023model}. This results in the fact that almost all applications are done in isolation from the wider community and make no attempt to support their validity through empirical observations. 

Empirical SwE is a field that has grown and become a mainstay in the broader field of SwE. Through that lens there are clear guidelines and methodologies that can be used to structure and perform empirical studies. Many of these approaches have been translated and re-used in the SyE context, such as the previously mentioned SLR \cite{cederbladh2023early}. Nonetheless, there is a large gap in the formalisation, benchmarking, and in general use of metrics in the SyE community \cite{henderson2023towards}. Without a more rigorous classification and collection of this kind of data, the possibility of performing empirical studies is greatly reduced. Here the SwE community can act as a baseline for how to improve this issue.

\begin{figure*}
    \centering
    \includegraphics[width=0.9\linewidth]{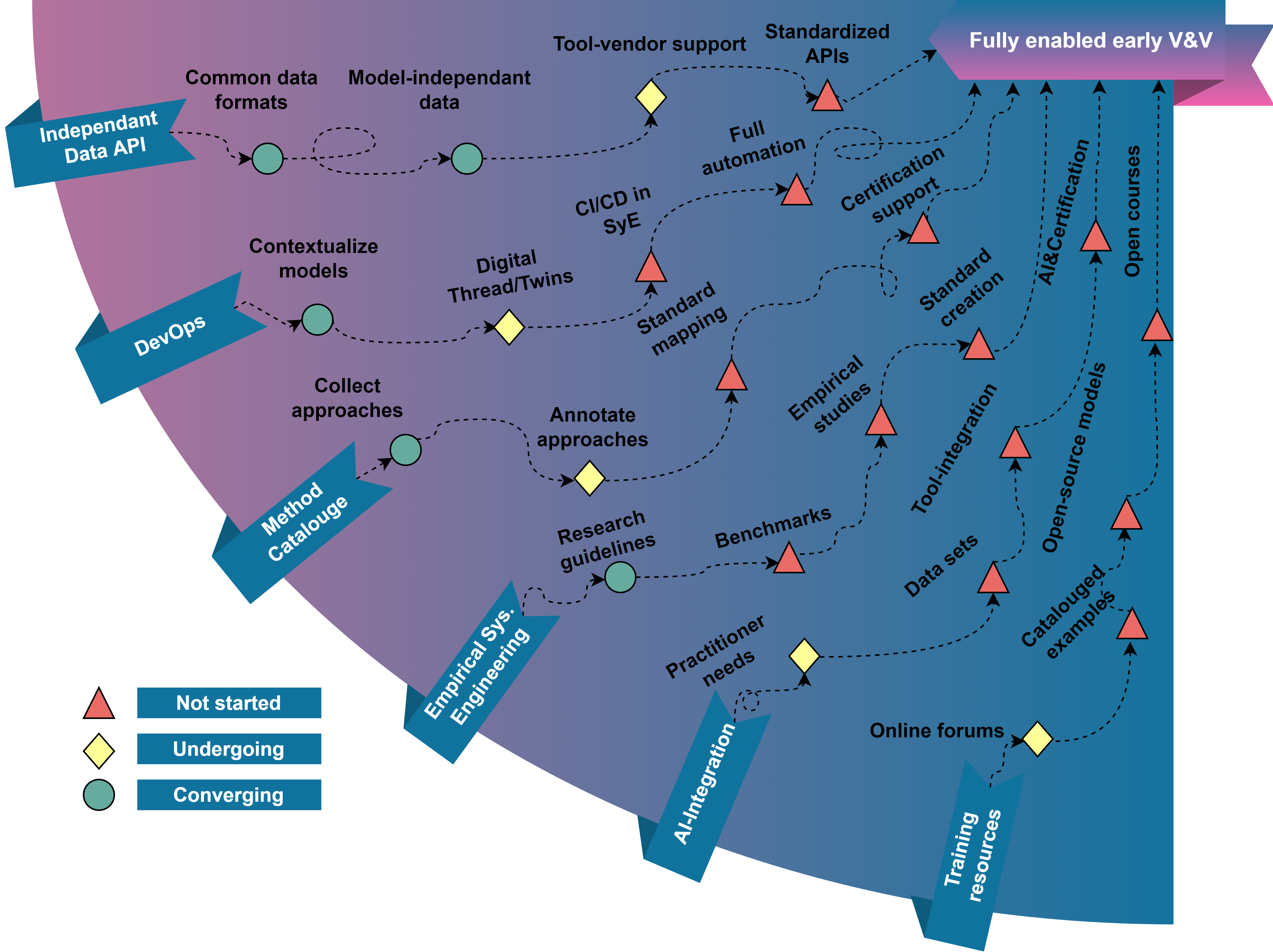}
    \caption{Different streams of research activities by researchers and practitioners to achieve early V\&V in MBSyE.
    }
    \label{fig:Roadmap}
\end{figure*}

\subsection{Tools (T1-T5)}

An often experienced challenge when adopting or learning model-based methods is that the available tooling is not satisfactory \cite{bucchiarone2020grand}. While this is not a novel challenge, it still remains and continues to inhibit wider adoption. Particularly in the MBSyE context there is a lack of open-source tools available to use, for example SysML tools are almost exclusively proprietary with licenses out of range for most individual users \cite{cederbladh2023early}. Additionally, most tools offer little to no support in cross-compatibility (even if the same modelling language is used) leading to a narrow vendor lock-in. As a result, there is often a huge divide in the academic contra industrial tools. In fact, academic solutions typically promote open-source or free tools so that outreach is greater, which is not a priority for companies. Another result is that most of the proposed novel academic solutions are not based or evaluated on real-life examples \cite{cederbladh2023early}, thus suffering of limited validation (notably in scalability). Modelling languages and tools should be interoperable in the wider company landscape. One important factor for this is that the model data itself needs to be made easier to integrate with and use, for example through open APIs. Without addressing these issues collaboration is reduced between different actors, notably academia and industry. The collaborative efforts of modelling itself can also be a challenge, notably to understand how tools should allow for multiple users to work together on the same model. 

The issue of modelling tools is something that has been largely investigated also in the MBSwE community. Indeed, tooling is one of the more commonly reported bottlenecks when it regards usability of modelling in industrial settings \cite{bucchiarone2020grand}. Many of the issues have translated directly to the world of MBSyE, but there are also additional issues concerning SwE tools used in the SyE context. Notably, the disparity between academic and industrial solutions is larger than compared to SwE \cite{cederbladh2023early}, and the use cases for tools are often holistic instead of specialised. 


\subsection{Community (C1-C4)}

The community involved in SyE and by extension MBSyE currently does not provide adequate resources for new practitioners to adopt and learn MBSyE skills \cite{de2022taxonomy}. Learning material that is available is often not sufficient for a novice user to learn the fundamental skills/tools expected from industry. Similarly, the MBSyE field is quite young and does not have strong foundations in research and education and as an extension, most actors in MBSyE are industrial \cite{cederbladh2023early}. Additionally, systems and software modelling in itself is often seen as a challenge by practitioners \cite{bucchiarone2020grand}. Adoption of model-based practices is a cross-cutting challenge as it can be reflected in many other challenges and not specific for early V\&V, although a more robust landscape for early V\&V could be an incentive to push user adoption. The audience is also very diverse, which means that it can be difficult to cater to all potential needs or expectations. There is also a notable lack of proper examples available online which can be used to train and learn. It is not feasible to expect that a practitioner can learn all necessary skills from a few examples, and central topics as systems thinking require more systematic learning. A more integrated SyE discipline in different master programs could be a good way to promote this \cite{de2022taxonomy}.

In SwE the issue of learning UML and similar notations have become a mainstay in education, and there exist many useful resources online with openly available examples and tutorials. The same cannot currently be said for MBSyE \cite{de2022taxonomy,henderson2021value}. For early V\&V, this results in a plethora of different disseminated approaches and techniques, that in reality cannot be re-used since there is too much implicit or hidden context. Essentially, there needs to be a bigger push from the community to embrace openly available artefacts and examples. The use of toy examples or other types of simplified examples could be a good first step towards solving this issue. 

\section{A research road-map} \label{Sec:Approach}


To address the challenges discussed so far while moving towards a more complete definition and understanding of early behaviour V\&V in MBSyE, we propose a high-level research road-map as depicted in Figure \ref{fig:Roadmap}. We discuss each thread of the road-map together with our motivation and reasoning. Notably, we consider how SwE research and methods can support the evolution of the field and advance the current V\&V technology and maturity in MBSyE. 

\subsection{Independent data API}

\textbf{Context.} Data is stored in models. It would be useful and beneficial for practitioners if it was possible to store this data in non-model specific means, so that it can be more easily re-used and applied.

\noindent
\textbf{Need.} Practitioners should interact with and interpret data in different models from different tools using standard APIs. This way, maintainability and organisational management of tools is greatly improved while promoting integrated solutions.

\noindent
\textbf{Gaps.} Currently, the way data is stored and interacted with is limited and there is a significant vendor lock-in. Due to this reason, it is quite difficult to manage the integration of various data sources and models in the overall landscape of MBSyE.

\noindent
\textbf{Software Engineering.} In SwE there is a longer history in this context, for example seen through the MDE context. Many of the general approaches and learning from the SwE domain should be transferable, probably requiring certain extensions due to the wider MBSyE context. 

\noindent
\textbf{Research} A first step is to use-common data formats, something that the industry is advancing towards for several aspects through the use of standards such as the System Structure and Parameterization (SSP) standard \cite{hallqvist2021engineering}. A part of this is to also include model-independent data in the often larger collection of models. This goal is pursued for example within the OSLC initiative \cite{basso2023model}, but still remains to be robustly integrated in industry. A next logical step is the tool-vendor support of various data standards in industry. This is of course a process that is always undergoing, but in our view we emphasise the need for tools to more readily support functions such as blended modelling and multi-model management/federation \cite{david2023blended}. A final step of this evolution would be the construction and usage of standardised APIs between models and their data with corresponding tools. Such a step however is dependent on a general industrial agreement to formulate the necessary pre-requisites. 

\subsection{DevOps}

\textbf{Context.} Often the notation used for system description is not the same as used for analysis, and providing seamless integration between notations in both directions is very useful.

\noindent
\textbf{Need.} For early V\&V to be successful, it should be performed in a \textit{continuous} way through the rest of development \cite{cederbladh2023early}. V\&V at the early stage is therefore not a ``one time thing'' and should propagate to further activities with potential feedback (using automation). 

\noindent
\textbf{Gaps.} Integrating DevOps in the wider SyE field is not straight-forward. There are many kinds of models and applications which need to be considered and fully integrated to provide a holistic coverage, and there are several distinct stages of development.

\noindent
\textbf{Software Engineering.} In the SwE domain paradigms like DevOps are well known \cite{ebert2016devops}. For SwE DevOps can already be considered as a standard practice in many industrial segments, and there are many technologies and techniques that can act as an inspiration if not directly be transferred to the MBSyE context. 

\noindent 
\textbf{Research.} A first step is to contextualise the various models found in the MBSyE process landscape. This is something that on a broader level already exists, but requires specific instantiation for specific organisations. Another necessary step is the creation and understanding of more general approaches for how the digital thread \cite{singh2018engineering} and digital twin \cite{jones2020characterising} paradigms can be integrated in the landscape. Currently these technologies do not have a strictly unified definition, although there are standards emerging \cite{ferko2022architecting}. From a more streamlined definition and understanding a move can be made to include more automated features, through the use of Continuous Integration (CI) / Continuous Delivery (CD) technologies \cite{ebert2016devops}. Integrating such pipelines effectively leads to the gradual shift towards more and more automation. Currently these technologies are missing, and it would be a valuable feedback to the early V\&V context through the use of already in place models and systems.

\subsection{Method catalogue}

\textbf{Context.} Currently the early V\&V landscape is ad hoc and there are a plethora of different (partially overlapping) activities used by practitioners that support early V\&V of system behaviour.

\noindent
\textbf{Need.} Practitioners and researchers need an accessible and well-founded catalogue of how various approaches, methods, tools, and techniques can support early V\&V of system behaviour and under what conditions and constraints they are valid.

\noindent
\textbf{Gaps.} Currently there are no general classifications and understanding of the concepts to provide the required catalogue. Notably, aspects of uncertainty in analysis validity is missing.

\noindent
\textbf{Software Engineering.} This issue is transdisciplinary across several domains, but there are examples of classifications in the domain of MBSwE which could be used as an inspiration \cite{gabmeyer2019feature}.

\noindent
\textbf{Research.} A first step would be to classify these approaches more explicitly to offer a detailed catalogue or classification for users, consisting of input/output relationships and conditions for usages; examples of such attempts can be found online\footnote{Multi-Paradigm Modelling catalogue: ttps://zenodo.org/record/2538711.Y0etVExBxaR} \cite{cederbladh2023early}. A next step is a classification pursuing improved understandability, as currently many methods are complex and not comprehensible for the average systems engineer. Advances need to be made towards explainability, something that is seeing increasing attention in SwE in relation to AI and other complex methods and approaches \cite{feldkamp2023explainable}. Through increased focus on explainability and transparency, the eventual followup is the mapping to various standards and processes. There needs to be a consolidated effort to understand what approaches can fulfil what purpose for certification in standards \cite{gotz2018verification}. Notably, what methods can be trusted to provide evidence or guarantees for what contexts, and what are the eventual conditions that apply? Eventually, methods should be applied for critical tasks such as certification with proof of result integrity and viability.

\subsection{Empirical Systems Engineering}

\textbf{Context.} To provide more substantial evidence for or against current early V\&V approaches empirical feedback is valuable.

\noindent
\textbf{Need.} More systematic studies adhering to well-defined standard practices to enable comparison and reasoning of the obtained effects.

\noindent
\textbf{Gaps.} Current early V\&V methods are almost exclusively demonstrated through examples which cannot be replicated, often with claims being made without any specific available data \cite{henderson2021value,cederbladh2023early}. Consequently, the results available in literature are often anecdotal or lack sufficient data for wider claims. 

\noindent
\textbf{Software Engineering. } In the SwE domain empirical studies have a strong foundation, and the issue of benchmarking solutions has been disseminated under a long time \cite{sim2003using}. There are also many commonly agreed upon metrics \cite{srinivasan2014software}. These include aspects such as uncertainty which is currently not well defined in MBSyE.

\noindent
\textbf{Research.} As a step towards mitigating this lack of empirical evaluation and structured feedback, there needs to be a consolidated effort to agree on what metrics should be used and how research should be conducted. There are works moving towards defining metrics \cite{henderson2023towards}, and these need to be further strengthened. A natural next step is the move towards defining benchmarks in the context of early V\&V, and how the (often) broader defined metrics for MBSyE can be applied. However, it must be noted that a part of the challenge of integrating various metrics and benchmarks for empirical studies is that MBSyE is heavily tied to the processes at hand. Therefore, most new techniques directly affect the process itself making it difficult to isolate/measure the possible effects and provide reproducible experiments. Nonetheless, leveraging metrics and benchmarks paves the way for more empirical studies to be performed and disseminated in industry. Eventually, there needs to be augmentations of existing standards and/or new proposals for standards to cover these topics.
This way, there is a way to formally define what are the acceptable practices to be used by practitioners and how to adopt various technologies in real projects.

\subsection{AI-Integration}

\textbf{Context.} Eventually, tooling should consider integrating various AI-based features that are seeing success in industry.

\noindent
\textbf{Need.} Tools as chat-bots and assistive technologies are great for aiding users, but also AI-based automation in terms of, e.g., evaluations to improve black-box analysis and general testing algorithms \cite{cabot2018cognifying}.

\noindent
\textbf{Gaps.} Currently the concrete capabilities AI can bring to the broader MBSyE context is not clear due to the availability of few examples, as well as a general lack of openly available data to be used.

\noindent
\textbf{Software Engineering.} In the SwE field AI is catching significant attention and provides a source of inspiration for several automation attempts. Additionally, in SwE the issue of certification in the context of generated artefacts has been extensively studied, and might act as a useful reference \cite{whalen2002synthesizing}. 

\noindent
\textbf{Research.} To include AI effectively there is first a need to understand \textit{what} AI can assist with in addition to \textit{how}. There needs to a more consolidated effort to get feedback on the needs of industrial practitioners and potential application areas. A parallel issue is that there is a need for data to integrate AI effectively, which is difficult to gather both in good quality and shared freely by companies. Without more robust datasets that can be used (preferably openly available) it becomes difficult to introduce AI technologies. Further, current industrial MBSyE tools offer little to no interface capabilities for AI technologies, so this needs to change in case the technology should be integrated more widely. Lastly, effective use of AI should not collide with certification of systems or their parts. This is however not feasible without much more effort into providing explainable AI solutions.

\subsection{Training resources}

\textbf{Context.} To promote best practices and formulate a common view of what early V\&V is and how it is applied there should be various training resources available.

\noindent
\textbf{Need.} Training resources should be easily accessible and provide clear and structured guides for users of various backgrounds. 

\noindent
\textbf{Gaps.} There are very few online training resources for MBSyE. For example the SysML, which is the de facto standard language, is quite difficult to train via online resources. Different V\&V methods in turn have typically even less training resources available. 

\noindent
\textbf{Software Engineering.} Modelling in the context of SwE has a larger set of resources available for practitioners, and a larger set of open-source or openly available examples. Often there is a partial overlap of uses for the modelling aspects for MBSyE, but for the eventual analysis there is a large gap between the domains.

\noindent
\textbf{Research.} It would be very valuable to create and maintain online forums for users of all backgrounds. There are few complete MBSyE model examples for any specific scope. A library or repository of examples can be used to guide and practitioners would be a good step forward. Eventually, there should be more efforts to create and maintain open-source models that can be used by anyone looking to learn the involved technologies, this would greatly improve the overall community integration. A final step to cement the training materials would be openly available and easily accessible training materials, which can cater to an as wide audience as possible.

\section{Related work} \label{Sec:RW}



This work aims to act as a reference for both industrial practitioners and researchers to work toward a common view in the context of MBSyE and early V\&V, one of the currently fragmented areas required to meet the expected future INCOSE 2035 SyE vision. Other works exist attempting a similar framing of parts of the SyE community. Henderson and Salado \cite{henderson2021value} provided a review of existing research to see what value MBSyE provided according to literature. Their work identified that around 90\% of works do not provide any empirical evidence for any claims of benefits observed in companies. Similarly, Campo \textit{et al.} \cite{campo2023model} perform a study on the value of MBSyE and their findings confirm what Henderson and Salado identified, with 86\% of authors not using any metric when discussing MBSyE value. Henderson \textit{et al.} have followed up the original work and proposed a first step towards metrics in Digital Engineering (DE), which encompasses MBSyE, in an attempt to mitigate the lack of metrics available in the field \cite{henderson2023towards}. Another paper by Carrol and Malins \cite{carroll2016systematic} investigates how MBSyE is justified. Their paper identifies a rapid interest in MBSyE but also a lack of concrete and factually reported benefits. The paper identifies a set of benefits, such as higher quality in requirements, design re-use, stronger V\&V, but reach a similar conclusion that the field has a fragmented view. 

Several gaps are identified by authors in the literature. De Saqui Sannes \textit{et al.} argue that MBSyE is being adopted by industry, but that tooling is still a major bottleneck \cite{de2022taxonomy}. Furthermore, the lack of strong educational foundations hampers the adoption. The lack and subsequent need for more robust education was also argued by Ramos \textit{et al.} \cite{ramos2011model}. Additionally the authors argued that the two most important challenges to overcome are to provide lean examples and methods in addition to more collaborative opportunities from the graphical modelling languages currently in use. Madni and Sievers performed a survey of MBSyE and elicited two large challenges in its wider adoption: the need for cultural change, and the capability to include non-modelling artefacts in the MBSyE methodology \cite{madni2018model}. Indeed, Huldt and Stenius also identified the challenge of cultural change together with the general management support as issues for MBSyE \cite{huldt2019state}. They also identify that the lack of perceived value of MBSyE is a large inhibitor, in addition to the overall learning curve and lack of industrial skill. Berschik \textit{et al.} report that while MBSyE is widespread and seeing increasing interest, there is a lack of common definitions and unified understanding of topics in the field \cite{berschik2023mbse}. Overall, the systematic studies in the context of MBSyE, while targeting different aspects, highlight many of the same issues we pinpoint in this paper. As such, our contribution takes a step further by proposing a corresponding research agenda which is based on what already exists in the SwE domain.


In the SwE community the notion of early V\&V for software behaviour relates back to the 80s \cite{wallace1989software}. As a result there are many technologies available to researchers and practitioners. However with technological advances new challenges emerge. More recently identified research opportunities include V\&V of AI behaviour \cite{zhang2020testing}, emphasising explainability of analysis. Similar challenges are identified for the autonomous vehicle domain, specifically for safety concerns \cite{rajabli2020software}. Souri \textit{et al.} also identify privacy and security as concerns currently in need of more research \cite{souri2020systematic}. At a glace some of these challenges comes from non-determinism from several software applications, but also due to the rise of specific attributes such as security. One aspect of note, is that several of the emerging domains in SwE considers a broader perspective than \textit{just} software. Indeed, as software becomes more central in systems often time systems and software engineers in the end work on the same concrete systems, for example in the form of CPS and autonomous vehicles. Therefore the cross-fertilisation of the fields is in some sense inevitable, but also necessary to tackle challenges of development.

\section{Conclusion} \label{Sec:Conclusion}

In this paper we have discussed the notion of early validation and verification in the context of Model-Based Systems Engineering. Notably, we extract the current state-of-the-art to highlight and provide a list of selected challenges. The challenges are summarised in six categories, \textit{Models}, \textit{Organisational}, \textit{Methods}, \textit{Scoping}, \textit{Tools}, and \textit{Community}. We discuss these challenges and then propose a research road-map for how to address these issues and improve the overall maturity of the practices related to early validation and verification. We consider six streams of research paths summarised as \textit{Independent data API}, \textit{DevOps}, \textit{Method catalogue}, \textit{Empirical Systems Engineering}, \textit{AI-integration}, and \textit{Training resources}.

The content of the paper is partially based on a Systematic Literature Review by the authors, and uses the review as the main input in conjunction with surrounding literature on the subject. 

\section{Acknowledgments}

This work was partly funded by the AIDOaRt project, an ECSEL Joint Undertaking (JU) under grant agreement No. 101007350.

\bibliographystyle{ACM-Reference-Format}
\bibliography{software}

\end{document}